\documentclass[aps,prb,twocolumn,superscriptaddress,nobibnotes]{revtex4-1}

\usepackage[T1]{fontenc}
\usepackage{bm}
\usepackage{multirow}
\usepackage{amssymb}
\usepackage{graphicx}
\usepackage{booktabs}
\usepackage{pifont}
\newcommand\cmark{\ding{51}}
\newcommand\xmark{\ding{55}}
\usepackage[colorlinks=true,linkcolor=blue,citecolor=blue,urlcolor=blue]{hyperref}
\pdfpageattr{/Group << /S /Transparency /I true /CS /DeviceRGB>>}

\begin{document}

\title{Engineering polar discontinuities in honeycomb lattices}

\author{Marco Gibertini}
\altaffiliation{These authors contributed equally to this work.}
\affiliation{Theory and Simulation of Materials (THEOS) and National Center for Computational Design and Discovery of Novel Materials (MARVEL), \'Ecole Polytechnique F\'ed\'erale
de Lausanne, CH-1015 Lausanne, Switzerland}

\author{Giovanni Pizzi}
\altaffiliation{These authors contributed equally to this work.}
\affiliation{Theory and Simulation of Materials (THEOS) and National Center for Computational Design and Discovery of Novel Materials (MARVEL), \'Ecole Polytechnique F\'ed\'erale
de Lausanne, CH-1015 Lausanne, Switzerland}

\author{Nicola Marzari}
\email{nicola.marzari@epfl.ch}
\affiliation{Theory and Simulation of Materials (THEOS) and National Center for Computational Design and Discovery of Novel Materials (MARVEL), \'Ecole Polytechnique F\'ed\'erale
de Lausanne, CH-1015 Lausanne, Switzerland}

\begin{abstract}
Unprecedented and fascinating phenomena have been recently observed at oxide interfaces between centrosymmetric cubic materials, such as LaAlO$_3$ and SrTiO$_3$, where a polar discontinuity across the boundary gives rise to polarization charges and electric fields that drive a metal-insulator transition, with the appearance of free carriers at the interface. Two-dimensional analogues of these systems are possible, and honeycomb lattices could offer a fertile playground, thanks to their versatility and the extensive on-going experimental efforts in graphene and related materials. Here we suggest different realistic pathways to engineer polar discontinuities across interfaces between honeycomb lattices, and support these suggestions with extensive first-principles calculations. Two broad approaches are discussed, that are based on (i) nanoribbons, where a polar discontinuity against the vacuum emerges, and (ii) selective functionalizations, where covalent ligands are used to engineer polar discontinuities by selective or total functionalization of the parent system. All the cases considered have the potential to deliver innovative applications in ultra-thin and flexible solar-energy devices and in micro- and nano-electronics. 
\end{abstract}

\maketitle

Combining together different materials rarely results in a simple
``arithmetic sum'' of their properties. Typically, the composite
system displays properties that are not present in its components,
giving rise to novel and unexpected behavior. This is the case 
for oxide interfaces,  which have been recently attracting
considerable attention, both  experimentally and theoretically\cite{Mann2010,Hwa2012}.
Among these, a dominant role is played by heterostructures of strontium
titanate (SrTiO$_{3}$ or STO) and lanthanum aluminate (LaAlO$_{3}$
or LAO). Both LAO and STO are insulators; nevertheless, when brought
together, a two-dimensional electron gas (2DEG) with high mobility
appears at their interface\cite{Oht2004}. This 2DEG is host to a
rich variety of phenomena, ranging from superconductivity\cite{Rey2007}
to magnetism\cite{Bri2007} (and even the unprecedented combination
of the two\cite{Li2011,Bert2011}), with many promising device applications. 
The most intuitive picture to explain the existence
of the 2DEG follows from dielectric considerations 
on the bulk properties of the constituent compounds\cite{Bris2014}.
LAO and STO have an identical cubic centrosymmetric crystal structure; 
therefore, classically, the macroscopic polarization of each material should be zero, thanks to inversion symmetry. 
However, in the framework of the Modern Theory of Polarization\cite{Resta1994}, 
polarization cannot be represented by a single vector, but as a lattice of vectors 
with the same periodicity of the crystal lattice and satisfying all the symmetries of the crystal. 
For cubic systems that display inversion symmetry, this condition allows two different polarization lattices, 
one containing the zero vector, and another shifted by half of the cube diagonal. 
STO belongs to the first class, while LAO to the second and, as a result, 
a polar discontinuity appears when LAO is epitaxially grown on top of STO 
and a polarization charge builds up at the interface\cite{Bris2014}. 
This discontinuity creates an electric field inside LAO that can in turn
induce a metal--insulator transition with a transfer
of free charges from the surface of LAO to the STO/LAO interface\cite{Bris2014,Jano2012}.
This transition has been found experimentally to occur at LAO thicknesses of 3-4 layers\cite{Thi2006},
in agreement with theoretical calculations\cite{Bris2014}.

Such concepts could be extended to lower dimensions, with one-dimensional (1D) 
channels of free carriers appearing at the boundary between two-dimensional (2D) insulating materials,
provided that the ``bulk'' polarizations of the 2D crystals involved were different. In this respect non-centrosymmetric
honeycomb lattices offer a very promising playground, owing to the quantized
and topological nature of their bulk polarization\cite{Fang2012,Jad2013}.
A suggestion in this direction has been put forward in
Ref.~\onlinecite{Bris2013}, where the authors have considered honeycomb crystals
of aluminum nitride (AlN), silicon carbide (SiC), and zinc oxide (ZnO).
First-principles simulations have confirmed the existence of a polar discontinuity 
at the interface between two such  crystals, 
with free charges accumulating in 1D channels along the interfaces\cite{Bris2013}. 
In the thermodynamic limit, the linear charge density $\lambda_{F}$ of free carriers 
perfectly balances the polarization charge density $\lambda_{P}$ and it is thus determined
solely by the bulk properties of the materials involved and by the
orientation of the interface through\cite{Vand1993} 

\begin{equation}
\lambda_{F}=(\bm{P}_{2}-\bm{P}_{1})\cdot\hat{\bm{n}}_{12}=-\lambda_{P}.\label{eq:1Ddensity}
\end{equation}
Here $\bm{P}_{1,2}$ are the bulk polarizations of the parent crystals
and $\hat{\bm{n}}_{12}$ is a unit vector normal to the interface
and pointing from 1 to 2. Although the suggestion for an analogue in 2D
of the LAO/STO heterostructure is very promising, a practical concern hinders
the feasibility of the setup suggested in Ref.~\onlinecite{Bris2013}. Indeed, although ZnO and SiC have been synthesized as few-layer hexagonal structures\cite{Tus2007,Lin2012}, they have not been isolated as monolayers, and the realization of in-plane heterostructures seems even more demanding.

\begin{figure*}
\includegraphics{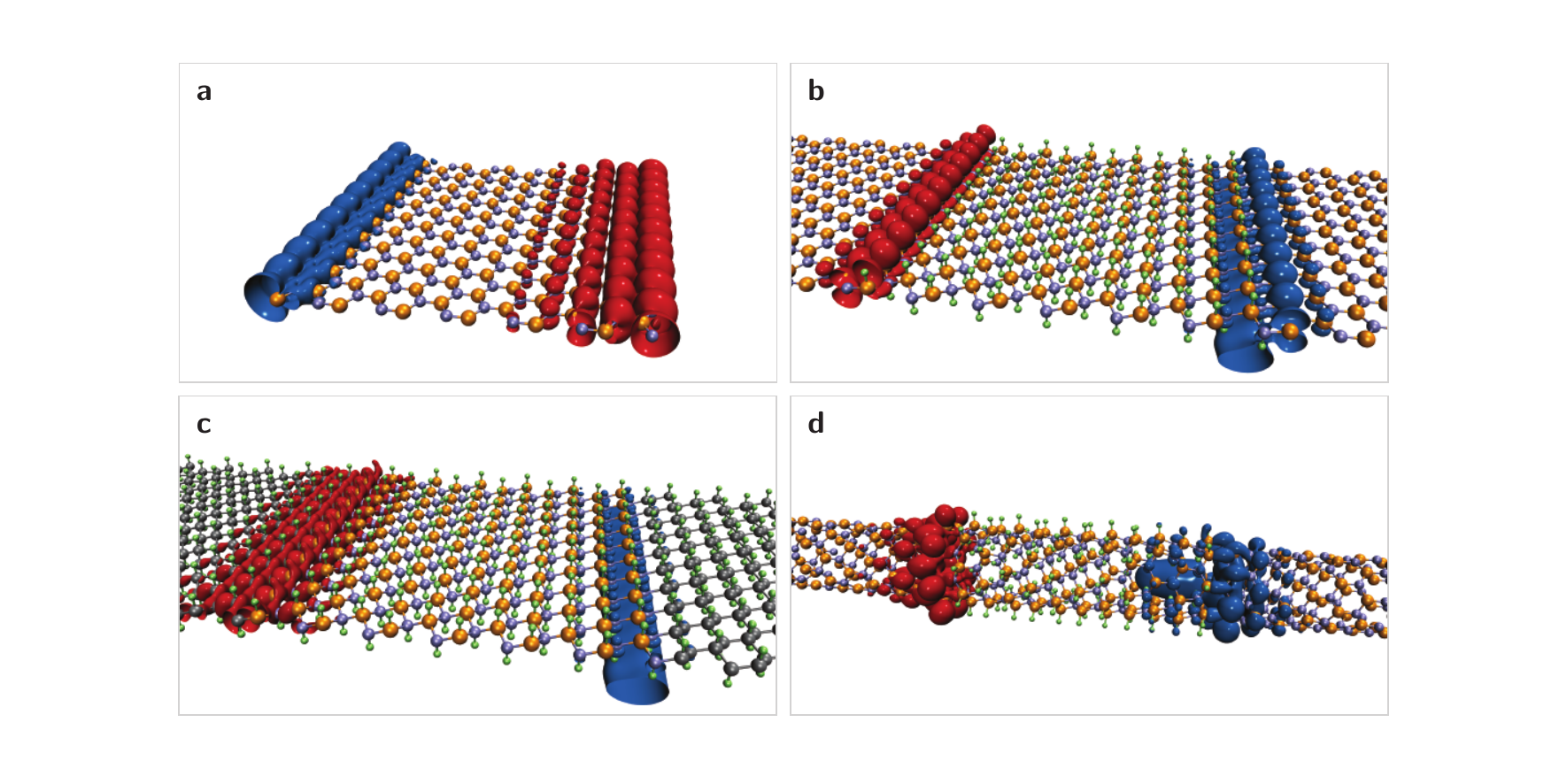}
\caption{\textbf{Possible realizations of a polar discontinuity in honeycomb
lattices}. \textbf{a}, finite-width nanoribbons cut out of a two-dimensional
heteronuclear honeycomb lattice (in this case boron nitride with unsaturated edges). The
nanoribbon is infinitely extended along one of the two in-plane directions,
while it has a finite width in the other direction. The interface
has to be considered with vacuum, an insulator with vanishing polarization.
\textbf{b}, interfaces arising from covalent functionalizations (e.g. with
hydrogen) of a parent honeycomb lattice (here, boron nitride). \textbf{c},
full functionalization of a heterostructure comprising graphene and
boron nitride, the two most extensively studied two-dimensional honeycomb
materials. \textbf{d}, nanotube obtained by rolling up a system like
the one in panel \textbf{b} in such a way that it still manifests
a polar discontinuity along its axis. In all panels, the
real space distribution of free-carrier density is also plotted in
red (blue) for electrons (holes). \label{fig:systems}}
\end{figure*}

In this paper we outline instead experimentally viable approaches to the realization of polar discontinuities
at the interface between honeycomb structures, 
giving rise to 1D wires of electrons and holes and opening the possibility of manifold applications.

First, we note that vacuum can be interpreted as an insulator
with vanishing polarization. Thus, in a nanoribbon made out of any polar lattice, polarization charges will
appear as a consequence of the polar discontinuity at its edges (i.e., at the interface with vacuum).
Second, we argue that covalent functionalizations (for instance with hydrogen
or fluorine) can change the polarization of the parent crystal. Partial
functionalization of a 2D sheet thus introduces a discontinuity in
the electric polarization at the boundary between functionalized and pristine regions. 
Alternatively, full functionalization of the recently reported\cite{Sut2012,Lev2012,Liu2013,Liu2014,Gong2014} lateral 
heterostructures between graphene and boron nitride can also be pursued. 
We illustrate in the following all these strategies in more detail, starting from those 
that are conceptually simpler and highlighting for each the challenge or simplicity
of their experimental realization. Last, we support these suggestions 
with the results of extensive first-principles numerical simulations.

\section*{Nanoribbons: \emph{sp} materials}

\begin{figure}
\includegraphics{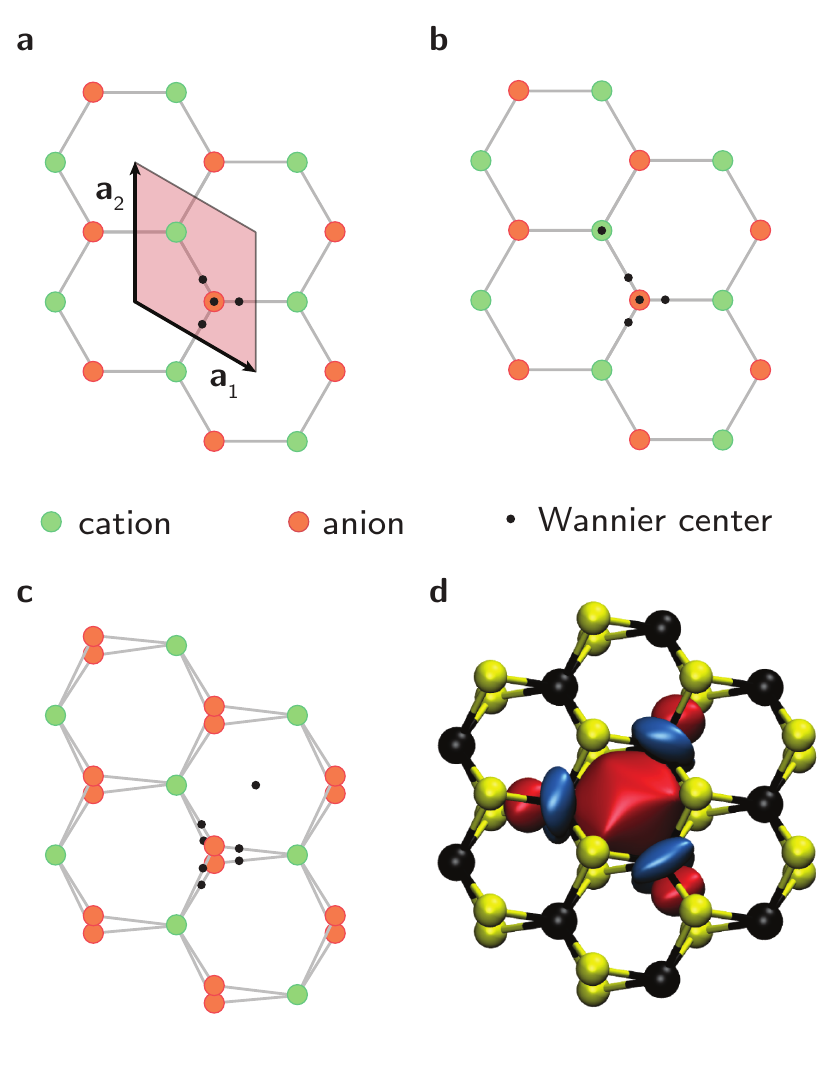}
\caption{\textbf{Maximally-localized Wannier functions in non-centrosymmetric honeycomb lattices}. \textbf{a},
crystal structure of a typical heteronuclear honeycomb lattice. Each
unit cell, identified by two lattice vectors $\bm{a}_{1}$ and $\bm{a}_{2}$,
includes two inequivalent lattice sites, occupied by a cation-like
and an anion-like atom, respectively. In \emph{sp} materials, the
upper valence bands can be mapped to four doubly-degenerate Wannier
functions centered around the anion. When the parent crystal is functionalized
with hydrogen in a chair configuration the effect is two-fold: (i) the
effective ionic charge at each lattice site is increased by one unit
and (ii) an additional Wannier function has to be included to accommodate
the extra electrons. As shown in \textbf{b}, the in-plane projection
of the corresponding center is on top of the cation while the other
four Wannier functions remain localized around the anion, as in panel
a. \textbf{c}, Wannier function centers for group-VI transition metal
dichalcogenides (e.g., MoS$_{2}$, WSe$_{2}$, etc.). Six doubly-degenerate
Wannier functions are located around the chalcongens while another
one is in the middle of the hexagonal cell. \textbf{d}, Isosurface
plot of the latter Wannier function in the case of MoS$_{2}$ (Mo
atoms in black and S atoms in yellow). \label{fig:wannier}}
\end{figure}

As a first suggestion we consider the case of pristine nanoribbons (see Fig.~\ref{fig:systems}a)
where the discontinuity has to be considered with vacuum. 
According to the interface theorem\cite{Vand1993}, the polarization
charge density is related to the bulk formal polarization\cite{Resta1994}
which, for non-centrosymmetric honeycomb crystals, is constrained
by symmetry to have quantized values and to point along one of the
equivalent armchair directions\cite{Fang2012,Jad2013,Bris2013}:

\begin{equation}
\bm{P}=\frac{e}{\Sigma}(\bm{a}_{1}+2\bm{a}_{2})\,\frac{m}{3}+\frac{2e}{\Sigma}\bm{R}\,.\label{eq:pol_honey}
\end{equation}
In equation~(\ref{eq:pol_honey}) $\bm{a}_{1}$ and $\bm{a}_{2}$
are the primitive lattice vectors (see Fig.~\ref{fig:wannier}),
$\bm{R}$ is a generic Bravais lattice vector, $\Sigma$ is the area
of a unit cell, and $m\in\{0,1,2\}$. The value of $m$ can be simply
obtained once the ground state of the system is expressed in terms
of a set of maximally-localized Wannier functions\cite{Mar2012}.
In addition, the apparent ambiguity in the choice of the lattice vector
$\bm{R}$ in equation (\ref{eq:pol_honey}) can be lifted by properly
assigning each Wannier function to a given ion according to the crystal
termination\cite{Ste2011}.
Then, the electronic contribution to $\bm{P}$ can be expressed as
a sum over point-like charges located at the Wannier centers $\langle\bm{r}\rangle_{j}$
and the total formal polarization reads%

\begin{equation}
\bm{P}=\frac{e}{\Sigma}\left(\sum_{\alpha=1}^{N}Z_{\alpha}\bm{\tau}_{\alpha}-2\sum_{j=1}^{N_{{\rm el}}/2}\langle\bm{r}\rangle_{j}\right)\,.\label{eq:pol_wannier}
\end{equation}
Here $Z_{\alpha}$ and $\bm{\tau}_{\alpha}$ are the charges and positions
of the $N$ ions in the unit cell and $N_{{\rm el}}$ is the number
of electrons. Let us first consider heteroatomic honeycomb lattices
in which the electronic properties are determined by \emph{s} and
\emph{p} orbitals, such as BN (monolayer SiC and ZnO, if realized, would also
belong to this class). In
Fig.~\ref{fig:wannier}a we show the Wannier function centers of
such systems and by using equation~(\ref{eq:pol_wannier}) it is easy to
show that the bulk formal polarization can be non-zero (see also Supplementary Information). 
A finite polarization charge thus
appears at the edges of a nanoribbon made out of one of these honeycomb
crystals, provided that the edge is not parallel to $\bm{P}$\cite{Vand1993}. 
We have verified that indeed the polarization charge vanishes for armchair nanoribbons
while it is maximal for perfect
zigzag nanoribbons. 

Looking at, e.g., the ideal case of zigzag edges, the polarization-induced 
electric field shifts the energy bands of the ribbon linearly in space, 
reducing the effective gap of the system. 
By increasing the width of the nanoribbon, a metal-insulator transition occurs 
as the top of the valence band reaches in energy the bottom of the conduction band. 
As shown in Fig.~\ref{fig:systems}a, free carriers localize close to the edges
of the nanoribbon with different character (electron or hole)
on opposite sides. The density of free carriers at each edge increases
with the width of the nanoribbon and, asymptotically, perfectly screens
the polarization charge in agreement with equation~(\ref{eq:1Ddensity}).
It is thus possible to tune such nanoribbons from a regime of small
widths (few nanometers) in which there is a sizable electric field
and negligible density of free carriers to an opposite regime for
large widths (tens of nanometers or more) of vanishing electric fields
but high metallicity. 

Metallicity of heteroatomic
zigzag nanoribbons has been investigated theoretically
in recent years\cite{Bar2008,Bot2008,Lou2011}, and 
it has been pointed out that such 1D metallic channels
can undergo magnetic transitions and eventually become half-metallic.
However, no connection with the intrinsic polarization of the parent materials
and with the existence of finite electric fields\cite{Qi2012} has been drawn yet. 
As we shall discuss later, this is
actually one of the key features that make some of these systems very promising
for solar-energy applications.

It is important to mention that, depending on the edge termination, additional charges
might appear at the boundaries. For instance, by terminating edge bonds with hydrogen, 
an additional $\pm e$ charge per unit length is introduced at opposite edges.
Therefore, any specific termination does not neutralize the total charge at the edge, since 
termination effects can change it only by an integer number of electrons per unit length, while 
the polarization charge is typically a non-integer fraction of an electron per unit length.
The only relevant exception are pristine III-V ribbons (i.e. exactly BN)
for which $m=0$, so that termination-induced charges might completely screen polarization effects.
On the other hand, functionalized BN ribbons, discussed later, would be resilient
against such mechanism. In addition, even in the case with $m=0$, by applying strain it would be possible to tune the polarization charge while leaving unaffected the termination effects, thus restoring a finite electric field (see also the Supplementary Information for a detailed description 
of termination effects). 

From an experimental point of view, the main challenge 
would be to control the edge structure and chemistry. 
On the other hand, recent progress\cite{Treier2011,Kim2013,Zha2013}
in atomistic control over the edge structure of graphene nanoribbons
has been quite spectacular and could be foreseeably extended to other honeycomb crystals. 

\section*{Nanoribbons: transition metal dichalcogenides}

In addition to \emph{sp} materials, other honeycomb lattices can support 
a finite bulk polarization, starting from transition metal dichalcogenides (MX$_{2}$). 
Although these materials have been extensively studied in the last few
years\cite{Wang2012,Chh2013}, their bulk formal polarization has not been discussed so far.
In such systems one sublattice is occupied by a transition
metal M while the other hosts two chalcogens X displaced in the vertical
direction on opposite sides with respect to the plane of M atoms.
In Fig.~\ref{fig:wannier}c we show the Wannier function centers
for the top seven valence bands when the transition metal belongs
to group VI (M=Mo, W). Six centers lie close to the
S atoms, which play the role of ``anions'' (see also Supplementary
Information), while the last one is located at the center of
the hexagonal cell and is associated with the Wannier function displayed
in Fig.~\ref{fig:wannier}d. As a consequence, group-VI transition
metal dichalcogenides like MoS$_{2}$ have a non-trivial (i.e. $m\neq0$)
formal polarization and their nanoribbons support a polar discontinuity,
in exact analogy with what happens for \emph{sp} materials. In addition, we mention that a polar discontinuity occurs also across inversion domain boundaries\cite{Zhou2013,Zan2013,Liu2014b} that lie along zigzag directions and separate crystallites with opposite polarizations. 
Thus, transition metal dichalcogenides offer a
broad choice in materials, chemistry, and electronic structure, and 
represent one of the most promising experimental avenues to pursue. 
Indeed, polarization effects might be at the origin of metallic states already observed at inversion domain boundaries in MoSe$_2$\cite{Liu2014b} and at the edge of MoS$_2$ nanoclusters\cite{Hel2010}.

\section*{Selective functionalization}

As a second different route to engineer polar discontinuities
we suggest covalent atomic functionalizations, such as those with hydrogen or fluorine. 
We assume full coverage and we consider for simplicity a chair conformation, corresponding
to functionalizations in alternating positions above and
below the plane of the parent honeycomb lattice
(see Supplementary Information for a discussion on the different conformations of functionalized BN and
their thermodynamic stability).
In Fig.~\ref{fig:wannier}b we show the Wannier function centers
for a typical functionalized honeycomb lattice. We report results
only for the case of hydrogen, since the case of fluorine is completely analogous.
It is easy to verify that these covalent functionalizations change the value of $m$
in equation~(\ref{eq:pol_honey}) by one unit with respect to the
parent material. As a consequence, we first note that functionalized nanoribbons 
would still support a polar discontinuity at the edges (and in 
particular III-V materials become  less sensitive 
to termination-induced charges as $m$ changes from
0 to 1, see Supplementary Table S1). 
Second, selective functionalization
of a parent honeycomb lattice would create an interface between pristine and functionalized
regions, introducing a polar discontinuity in the system. This
situation is depicted in Fig.~\ref{fig:systems}b, where we consider
perfect zigzag interfaces 
(giving rise to the largest polar discontinuity)
between alternating stripes of pure and hydrogenated BN, similarly to what happens in graphene ``nanoroads''\cite{Singh2009}.
As expected, free carriers localize at opposite interfaces. 

This case  will be discussed at length in the last part of the paper, but we point out 
that the two regions (pristine and functionalized) remain obviously
aligned with respect to each other and  are close to a perfect 
lattice match (e.g., for BN: $a_{\rm BN}=2.51$~\AA\,  and $a_{\rm BNH_2} = 2.59$~\AA). 
The experimental challenge is thus shifted to the selective functionalization of the parent crystal.
In order to achieve this result, it is likely that techniques adopted
for the functionalization of graphene\cite{Kar2013,Joh2013} could
be generalized to heteroatomic honeycomb crystals. 
Indeed, full coverage, double-sided hydrogenation of graphene
(i.e.~graphane) has been realized in suspended samples by exposure 
to low-temperature hydrogen plasmas\cite{Eli2009}. As far as fluorographene is concerned, 
a 1:1 carbon to fluorine ratio is achievable by functionalization with atomic fluorine 
formed by decomposition of xenon difluoride  (XeF$_{2}$)\cite{Nair2010,Rob2010}. 
By combining this technique with scanning probe lithography a pristine graphene nanoribbon 
has been isolated within a matrix of partially fluorinated graphene\cite{Lee2011}. 
In addition, encouraging results have been already reported on the partial fluorination 
of BN nanotubes\cite{Tang2005} and  nanosheets\cite{Xue2013}.

\begin{table*}
\caption{\textbf{Methods and materials to engineer polar discontinuities}
Possible methods and materials to engineer a 
polar discontinuity in honeycomb lattices. The list of materials 
could be greatly enlarged, but we have intentionally restricted it 
to current experimentally relevant materials.
We remark that the polar discontinuity is largest for zigzag interfaces or edges, 
while it is zero for armchair ones.
\label{tab:summary}}
\centering
\begin{tabular}{r p{0.1cm} p{9cm}}
\toprule
{\bfseries Methods}					&&    {\bfseries Materials} \\
\midrule
\multirow{2}{*}{Nanoribbons}	&&    BN, functionalized BN, \\
								&& transition metal dichalcogenides (MoS$_2$, \dots) \\ 
Inversion domain boundaries 	&&	 transition metal dichalcogenides \\
Selective functionalization	&&   BN \\
Full functionalization			&&	graphene/BN  heterostructures\\
\bottomrule 
\end{tabular}
\end{table*}

\section*{Functionalized graphene/boron nitride interfaces}

In view of the well-established experimental technology in growing single layer graphene
and boron nitride and the recent achievements\cite{Sut2012,Lev2012,Liu2013,Liu2014,Gong2014} 
in obtaining sharp graphene/BN lateral heterostructures, it is of great interest to exploit
these materials to engineer a polar discontinuity. 
While pristine graphene is not an insulator and does not support
a bulk polarization, its functionalized forms (graphane\cite{Eli2009}
and fluoro-graphene\cite{Nair2010,Rob2010}) are insulators and their formal polarization
is constrained by symmetry to be zero\cite{Fang2012,Jad2013}.
Moreover, we have seen above that functionalized BN acquires a
non-trivial bulk polarization ($m=1$). Thus, full functionalization
of existing planar graphene/BN heterostructures\cite{Sut2012,Lev2012,Liu2013,Liu2014,Gong2014} 
will lead to the emergence of a polar discontinuity and a finite
density of free carriers at the interfaces, as shown in Fig.~\ref{fig:systems}c
(we stress that this mechanism is completely
different from the one that leads to metallicity in unfunctionalized
graphene/BN interfaces\cite{Pru2010}). 
In addition, the intrinsic preference of these interfaces to grow along a zigzag direction\cite{Sut2012,Liu2014} provides the optimal orientation to maximize the polar discontinuitiy.

\section*{Functionalized nanotubes}

Although this paper is mainly devoted to 2D systems, we would like
to illustrate what happens when these honeycomb lattices are rolled
up in nanotubes and a finite polarization along the axis arises depending 
on the chirality of the nanotube\cite{Nak2003}. Focusing on zigzag
nanotubes, by selective functionalization one can introduce a polar discontinuity along the tube, 
as illustrated in Fig.~\ref{fig:systems}d. 
Similarly to what happens in 2D, a finite 
polarization charge builds up at the interfaces, creating an electric
field that induces a charge reconstruction, with the appearance of electron- and 
hole-rich quantum dots.
The charge density localized in these quantum dots is shown in Fig.~\ref{fig:systems}d
in the case of a (8,0) BN nanotube with selective hydrogen functionalization. 
The reduced dimensionality suggests that the effects of Coulomb interactions
might be relevant for the electronic-structure properties of such
quantum dots, similarly to what happens in carbon-nanotube quantum dots\cite{Sap2006}. 
The interaction-driven phenomena that might arise would
then be interesting both from a fundamental and practical point of
view, with particular emphasis towards quantum information applications\cite{Los1998}. 
In addition, even in the regime of small system sizes (when no charge
is transferred in the quantum dots), the magnitude of the electric
field in each segment might be easily tuned by varying
the diameter of the nanotube and the distance between the interfaces.
As we shall discuss in the following, this has significant consequences
in solar-energy applications.

\section*{Results and discussion}

In order to support the general arguments presented above,
we now use detailed, large-scale first-principles simulations to investigate
a paradigmatic case study. For definiteness we focus on selective
hydrogen functionalization of BN (that we label as BNH$_{2}$), even though qualitatively similar
results can be obtained using different parent materials and functional
atoms or any of the alternative approaches discussed above (and summarized
in table~\ref{tab:summary}). 

\begin{figure*}
\includegraphics{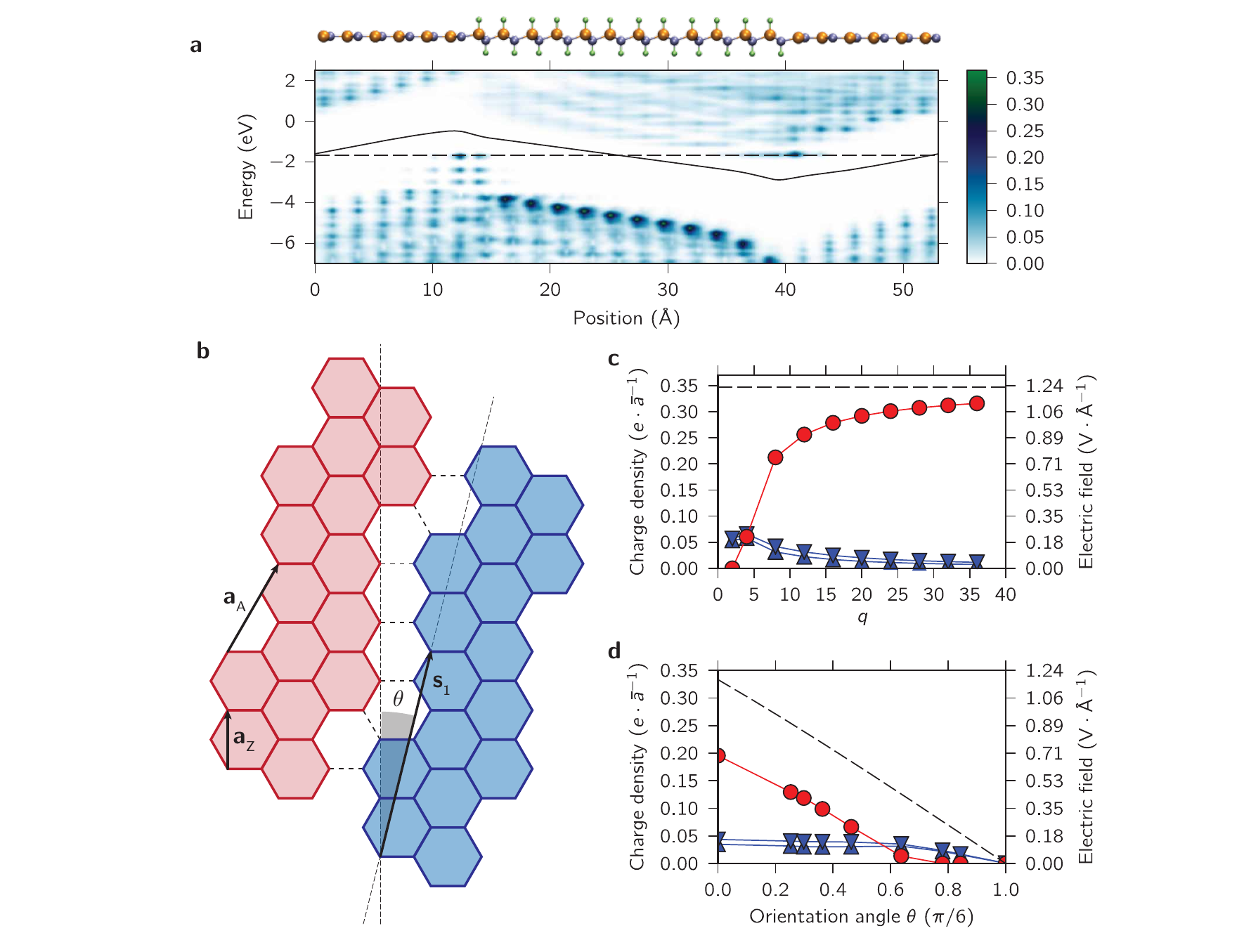}
\caption{\textbf{Numerical results for BN-BNH$_{\bm{2}}$ interfaces}. \textbf{a},
planar average of the local density of
states as a function of energy and position along an axis orthogonal
to the interfaces, for a $q=12$
BN-BNH$_{2}$ superlattice with $p=0$ (zigzag interface) shown
in lateral view on top. The dashed line marks the Fermi
energy while the solid line denotes the average
electrostatic potential energy. \textbf{b}, schematic representation
of an arbitrary interface orientation. The lattice vector along the
interface, $\bm{s}_{1}=\ell \bm{a}_{Z} + p\bm{a}_{A}$, forms an angle
$\theta$ with a zigzag direction; $\ell=2$ and $p=1$ here.
\textbf{c}, free charge density (red circles) accumulated at each zigzag interface
as a function of the superlattice periodicity. Charge is 
expressed in units of $e/\bar{a}$, with $\bar{a}$ being the equilibrium 
lattice constant along the interface. The horizontal
line shows the expected asymptotic value given by equation~(\ref{eq:1Ddensity}).
Up (down) triangles denote the residual average electric fields inside
BN (BNH$_{2}$). \textbf{d}, same as in panel \textbf{c} but as a
function of the interface orientation angle $\theta$ for a $q=7$
superlattice. The dashed line shows the free charge
density in the limit $q\to\infty$, equation~(\ref{eq:pol_orientation}).
In panel \textbf{c} and \textbf{d} the scales of left and right axes
are chosen to map into each other by proper
geometrical and physical factors. \label{fig:numerics}}
\end{figure*}

We simulate interfaces between pristine and functionalized BN 
within periodic boundary conditions, considering superlattices
obtained by alternating BN and BNH$_{2}$ regions. As a consequence,
two opposite interfaces are present within a simulation supercell,
which is identified by two primitive lattice vectors. 
The first one defines the periodicity along the interface and can be determined by specifying
the number of zigzag and armchair sections present, as shown in Fig.~\ref{fig:numerics}b:
$\bm{s}_{1}=\ell\bm{a}_{{\rm Z}}+p\bm{a}_{{\rm A}}$ ($\ell$ and $p$ are 
positive coprime integers while $\bm{a}_{{\rm Z}}=\bm{a}_{2}$
and $\bm{a}_{{\rm A}}=\bm{a}_{1}+2\bm{a}_{2}$ are translation vectors
along the zigzag and armchair direction, respectively). 
The second lattice vector ${\bm s}_{2}=2q {\bm a}_{1}$ defines the periodicity of the superlattice
in terms of the number of unit-cell repetitions $q$ that define the width of each region.
It is important to mention that the lattice vector $\bm{s}_{1}$ alone is not sufficient to uniquely define
the interface, since one still needs to specify the shape of the boundary (i.e. which lattice sites should be
assigned to each side of the interface). For simplicity we consider only
interfaces that minimize the number of boundary atoms and bonds\cite{Akhm2008}. 
Different choices would affect only quantitatively the results, 
through the appearance of additional bound charges at the interface. 

We first focus on the case of a perfect zigzag interface ($p=0$), when
we have a finite polarization charge density with opposite signs at each interface. 
These polarization charges create an electric field inside both BN and BNH$_{2}$, as can be clearly seen
in Fig.~\ref{fig:numerics}a examining the finite slope of the macroscopic
and planar average\cite{Bal1988} of the electrostatic potential energy (solid line). In
Fig.~\ref{fig:numerics} we also show the average local density of
states (LDOS) as a function of energy and position along the direction 
orthogonal to the interfaces. 
As a consequence of the electric fields, 
the electronic bands shift linearly as a function of position. For sufficiently large
widths (as in Fig.~\ref{fig:numerics}a), this leads
to an energy overlap between the conduction and valence bands of the
two interfaces, and to a charge redistribution with the
creation of electron and hole pockets\cite{Jano2012,Bris2013}. The
Fermi energy (dashed line) intersects both the top of the valence band and
the bottom of the conduction band so that the system has become metallic, as expected.
Fig.~\ref{fig:systems}b shows the excess charge density obtained
by  integrating the LDOS in order to take into account the
partial depletion of the valence bands (for holes) and the filling of
the conduction bands (for electrons). Both figures make it clear
that excess electrons and holes are separated in space and reside
on opposite interfaces, partially screening the polarization charges.
In Fig.~\ref{fig:numerics}c we report the density of free carriers
for different widths of the ribbons (red circles). As the periodicity
of the superlattice increases, the charge of free electrons and holes
also increases as a result of a larger overlap between conduction
and valence bands. Since the free charge has an opposite sign with
respect to the polarization charge, the overall charge density at
each interface decreases together with the electric field in both
materials (blue triangles in Fig.~\ref{fig:numerics}c). Asymptotically,
the free charge completely balances the polarization charge according
to equation~(\ref{eq:1Ddensity}) and the electric fields vanish,
thus preventing a \emph{polar catastrophe}\cite{Bris2014}. Indeed,
in Fig.~\ref{fig:numerics}c the density of free carriers approaches
asymptotically the polarization charge
obtained from the bulk formal polarizations of BN and BNH$_{2}$ (black solid line).
If the two materials were perfectly lattice-matched we would have,
from the discussion on Wannier functions above [equations~(\ref{eq:1Ddensity})
and (\ref{eq:pol_honey})], that $|\lambda_{P}|=e/(3a)$. Owing to
the piezoelectric properties of these materials, $\lambda_{P}$ is slightly
larger than $e/(3\bar{a})$ as a result of the finite strain necessary
to reach a common equilibrium lattice constant $\bar{a}$ along the interface\cite{Bris2013}.

Let us now consider an arbitrary interface orientation that can be
identified by the angle $\theta$ between the lattice vector along
the interface, $\bm{s}_{1}$, and the pure zigzag direction, $\bm{a}_{Z}$,
so that 
\begin{equation}
\cos\theta=\frac{\bm{s}_{1}\cdot\bm{a}_{Z}}{|\bm{s}_{1}||\bm{a}_{Z}|}=\frac{3p+2\ell}{2\sqrt{3p^{2}+3p\ell+\ell^{2}}}\,.
\end{equation}
 According to equations~(\ref{eq:1Ddensity}) and (\ref{eq:pol_honey})
the polarization charge density gradually decreases down to zero as
$\theta$ goes from zero (pure zigzag, $p=0$) to $\pi/6$ (pure armchair,
$\ell=0$). In particular, neglecting for simplicity piezoelectric
effects, we find 
\begin{equation}
|\lambda_{P}|=\frac{2e}{3\bar{a}}\sin\left(\frac{\pi}{6}-\theta\right)\,.\label{eq:pol_orientation}
\end{equation}
Thus, we expect that the appearance of finite electric fields and
the presence of a metal-insulator transition are not restricted to
the case $\theta=0$, although their effects are depressed as we approach
$\theta=\pi/6$. Fig.~\ref{fig:numerics}d shows the free charge
density and the electric fields in BN and BNH$_{2}$ for several values
of $\theta$ corresponding to different combinations of $\ell$ and
$p$. All simulations have been performed keeping fixed the periodicity
of the superlattice ($q=7$). 
Despite the small width of the system, the free charge density 
survives over a wide range of angles, thus suggesting robustness with respect to
the interface orientation. In addition, by incrementing the periodicity of
the superlattice the free charge density could be further increased
and asymptotically reach the solid line representing the polarization
charge in equation~(\ref{eq:pol_orientation}), as it happens in
the $\theta=0$ case shown in Fig.~\ref{fig:numerics}c.

\section*{Conclusions}

We have presented different approaches to obtain polar
discontinuities in honeycomb lattices, supporting these predictions
with first-principles simulations. First, we highlight that
a finite-width nanoribbon introduces a polar discontinuity with vacuum
if the parent material supports a finite formal polarization. This
happens for heteroatomic honeycomb crystals such as boron
nitride and its functionalized derivatives, and for transition metal dichalcogenides, such as molybdenum disulfide. 
The existence of a polar discontinuity at the edges elucidates why 
metallicity can arise in honeycomb nanoribbons\cite{Bar2008,Bot2008,Lou2011} or at inversion domain boundaries\cite{Liu2014b}. 
Second, we show that covalent atomic functionalizations, e.g. with hydrogen or fluorine,
can change the bulk polarization of a honeycomb lattice. Thus, covalent
functionalizations can be used to engineer polar discontinuities 
in 2D materials or 1D nanotubes simply by introducing interfaces between
functionalized and pristine sections. In addition, since covalent
functionalizations open a gap in graphene, they can be exploited
to engineer polar discontinuities in existing graphene/BN interfaces\cite{Sut2012,Lev2012,Liu2013,Liu2014,Gong2014},
without the need for selective functionalization. 

We believe that engineering polar discontinuities in honeycomb lattices will
provide a novel platform for manifold applications. First,
1D channels of free carriers along the interfaces could be exploited
for circuitry in new-generation ultra-thin and flexible electronics.
Indeed, current signals between different units of a device could
be transmitted along such 1D channels, surrounded by insulating bulk
materials, exceeding the limits of lithography in current electronic
devices. Moreover, the reduced dimensionality of the channels gives
rise to magnetic instabilities\cite{Bris2013} that could be useful
in spintronics applications. Second, and even more compelling,
we envision a fruitful employment in solar-energy technology 
for the realization of light-harvesting devices. 
Indeed, the ``bulk'' interior of these systems is insulating and is
an active region where photons can be absorbed, creating electron-hole
pairs. For narrow systems, the polarization charges at the interfaces or edges
are not compensated and thus naturally create an electric field 
(we note in passing that this is different from what has been done in  recent photovoltaic devices based on transition metal dichalcogenides\cite{Posp2014,Bau2014,Ross2014,Jo2014}, 
where metallic gates have been employed to create a $p$-$n$ junction). Once the electron-hole
pair is created, the electric field separates the electron and the hole and guides them
towards opposite interfaces, where the 1D wires naturally collect and transport them. 
In addition, the electric field shifts in space the conduction and valence band
extrema, and creates a variable effective gap depending on the spatial extension
of the exciton, with an ensuing tunability  of the cell efficiency. 
Furthermore, several systems with different widths and materials composition
could be integrated into a single device in order to optimize the
range of photon frequencies that can be absorbed.

\section*{Methods}

All first-principles calculations reported here are carried out within
density-functional theory (DFT) by using the PWscf code of the Quantum-ESPRESSO
distribution\cite{Gian2009} with the Perdew-Burke-Ernzerhof exchange-correlation
functional\cite{PBE}. 
An ultrasoft pseudopotential description\cite{Vand1990} of the ion-electron
interactions is adopted. Energy cutoffs are set to 60~Ry and 300~Ry
respectively for the electronic wavefunctions and the charge density
in the case of BN/BNH$_{2}$ superlattices. For zigzag interfaces
a $1\times6\times1$ shifted Monkhorst-Pack grid is
used to sample the Brillouin zone together with a $0.01$~Ry Marzari-Vanderbilt
smearing\cite{Marz1999}. 
In order to simulate a 2D system irrespective
of the three-dimensional periodicity requirements of plane-wave basis
sets, a vacuum layer of 20~\AA~ is added between periodic replicas in
the vertical direction. Relaxed structures are obtained within the
Broyden-Fletcher-Goldfarb-Shanno method by requiring that the forces
acting on atoms are below 0.026~eV/\AA~and the residual stress on the
cell is less than 0.5~kbar. Some simulations have been performed
without relaxation in order to simplify the calculations without qualitatively
affecting our results. 
We notice that the well-known DFT limitations in predicting energy gaps influence only quantitatively the relation between free charge density, electric fields, and width of the system, without changing the general physical picture or its asymptotic limits.
Maximally-localized Wannier functions have
been computed using Wannier90\cite{Wannier90}. 
We created figures of structures and charge densities using VMD\cite{VMD}.

\section*{Acknowledgements}
Simulation time was provided by the Swiss National Supercomputing Centre (CSCS) through project ID s337; 
M.G. acknowledges partial support by the Max Planck--EPFL Center for
Molecular Nanoscience and Technology. This research was stimulated by
a talk of E. Artacho on the work of Ref.~\onlinecite{Bris2013}.

\section*{Author contributions}
M.G., G.P. and N.M. conceived the work; 
M.G. and G.P. performed the first-principles simulations, and 
M.G., G.P. and N.M. wrote the manuscript.

\section*{Competing financial interests}
The authors declare no competing financial interests. 

\renewcommand\thefigure{S\arabic{figure}}
\setcounter{figure}{0}
\renewcommand\theequation{S\arabic{equation}}
\setcounter{equation}{0}
\renewcommand\thetable{S\Roman{table}}
\setcounter{table}{0}

\section*{Supplementary Information}

\subsection*{Effects of termination in nanoribbons}

\begin{figure}
\includegraphics{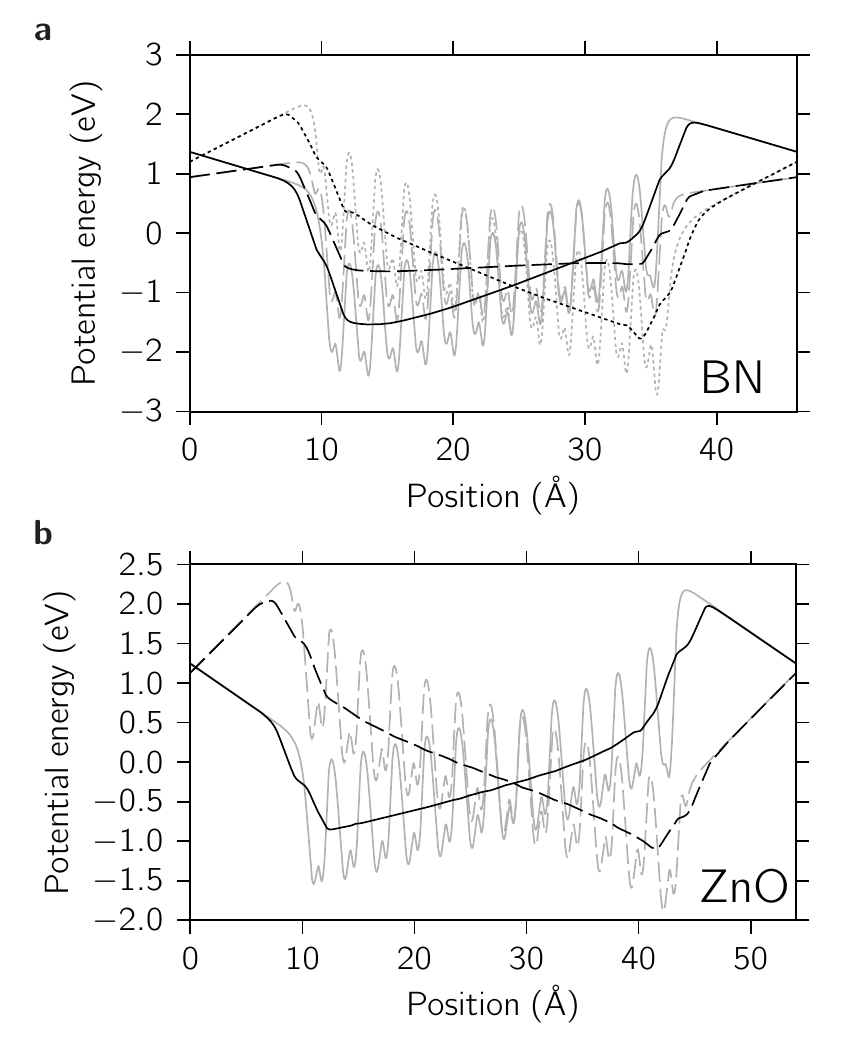}
\caption{\textbf{Electrostatic potential energy in nanoribbons}. Macroscopic
average (black lines) on top of the planar average (gray lines) of
the electrostatic potential energy as a function of the in-plane coordinate
orthogonal to the nanoribbon edges. Results are reported for boron nitride (BN, \textbf{a}) 
and a theoretically postulated zinc oxide 2D hexagonal crystal (ZnO, \textbf{b}). 
Solid and dashed lines refer to ribbons with unpassivated and hydrogen-terminated
dangling bonds respectively. An electric field is present in materials
like ZnO, which have $m\neq0$ in equation~(\ref{eq:pol_honey}), 
irrespective of the edge termination. 
In panel {\bf a} results for the case of double-hydrogen passivation 
are also reported (dotted lines).\label{fig:ribbon-potentials}}
\end{figure}

Whenever a honeycomb crystal supports a finite formal polarization,
a polar discontinuity arises at the interface with vacuum, as for instance 
in finite-width nanoribbons. We thus expect a finite electric field to be present as
a result of the polarization charge density $\lambda_{P}$ that appears
at the edges. This electric field will in turn trigger a metal--insulator
transition with increasing width of the nanoribbon. On the other hand,
depending on the specific termination of the nanoribbon, we may have
additional bound charges at the edges that partially screen the polarization
charge. These contribute to the total bound charge density $\lambda_{E}$
at each edge with a integer multiple of $e$ per unit length. According
to equation~(2), the total bound charge density at the
edge for a zigzag nanoribbon reads

\begin{equation}\label{eq:lambdaE}
\lambda_{E}=\frac{e}{a}\left(\frac{m}{3}+n\right)\,.
\end{equation}
As we mentioned in the paper, the integer $m\in\{0,1,2\}$ is completely
determined by the bulk properties of the system, while the integer
$n$ depends on the specific termination of the nanoribbon. We thus
have that whenever $m=0$, the total charge at the edge (and consequently also the
corresponding electric field) might vanish depending on the nanoribbon
termination. To show this,  in Fig.~\ref{fig:ribbon-potentials}a
we plot the macroscopic average (black lines) on top of the planar average 
(gray lines) of the electrostatic potential energy across a BN nanoribbon 
($m=0$, see table~\ref{tab:m-values}), including also a region of vacuum on both sides of the ribbon. 
We see that an electric field is present when edge bonds are unpassivated (solid lines, $\lambda_{E}=-e/a$,
i.e. $n=-1$) while it disappears when bonds are saturated with hydrogen
(dashed lines, $\lambda_{E}=0$, i.e. $n=0$).
A finite electric field is restored (although with opposite sign) when dangling bonds of edge atoms are saturated with two hydrogen atoms (dotted lines, $\lambda_{E}=e/a$, i.e. $n=1$).

On the contrary, when $m\neq0$ a non-vanishing bound charge (and the
corresponding electric field) will always be present at the edges
irrespective of the termination, although it may change sign depending
on the value of $n$. Indeed, the polarization charge in these cases
is \emph{fractional} and thus can not be completely screened by an
integral termination-induced edge charge. This is shown in Fig.~\ref{fig:ribbon-potentials}b in the case of ZnO 
($m=1$, see table~\ref{tab:m-values}). The electric field associated with the polarization
charges is always different from zero, although it changes
sign as the polarization charge goes from $\lambda_{E}=-2e/(3a)$
(i.e. $m=1$, $n=-1$) for unpassivated edge bonds (solid lines) to $\lambda_{E}=e/(3a)$
(i.e. $m=1$, $n=0$) for hydrogen-terminated bonds (dashed lines). 
We mention that similar conclusions apply also for functionalized BN nanoribbons ($m=1$, see table~\ref{tab:m-values}).

\begin{table}
\caption{Value of $m$ in equation~(\ref{eq:pol_honey}) for several material classes, 
together with their possible applications to obtain
a polar discontinuity. Some materials have been chosen as representative
examples for each class: BN (III-V); MoS$_{2}$ (transition metal dichalcogenides);
SiC (IV-IV); ZnO (II-VI), even if SiC and ZnO have not been synthesized in a 2D honeycomb lattice 
(for this reason we separate these to the second half of the table). BNH$_{2}$ stands for functionalized
III-V materials and similarly for SiCH$_{2}$ and ZnOH$_{2}$.\label{tab:m-values}}

{\centering
\begin{tabular}{cccc}
\toprule
Material class & $m$ & Nanoribbon & Selective functionalization\tabularnewline
\midrule
BN & 0 & \xmark\footnotemark[1] & \multirow{2}{*}{\cmark\phantom{\footnotemark[2]}}\tabularnewline
BNH$_{2}$ & 1 & \cmark\phantom{\footnotemark[1]} & \tabularnewline
\midrule 
MoS$_{2}$ & 1 & \cmark\phantom{\footnotemark[1]} & \cmark\footnotemark[2] \tabularnewline
\midrule\midrule
SiC & 2 & \cmark\phantom{\footnotemark[1]} & \multirow{2}{*}{\cmark\phantom{\footnotemark[2]}}\tabularnewline
SiCH$_{2}$ & 0 & \xmark\footnotemark[1] & \tabularnewline
\midrule
ZnO & 1 & \cmark\phantom{\footnotemark[1]} & \multirow{2}{*}{\cmark\phantom{\footnotemark[2]}}\tabularnewline
ZnOH$_{2}$ & 2 & \cmark\phantom{\footnotemark[1]} & \tabularnewline
\bottomrule 
\end{tabular}}

\raggedright\small $^1$ In this case a polar discontinuity is still possible but some
nanoribbon edge terminations can remove it.

\small $^2$ In principle selective functionalization works for transition
metal dichalcogenides, even though it would be difficult to realize
in practice.
\end{table}

To summarize, in nanoribbons made of materials for which $m\neq0$
in equation~(\ref{eq:pol_honey}) the polarization charge can not
be completely screened irrespective of the edge termination and thus
a finite electric field is always present, which induces a metallic
state for sufficiently large widths. Nanoribbons with $m=0$ might
still support a finite electric field but would be more susceptible
to termination-induced effects. Finally, we stress that while the termination-induced bound charge can be obtained by charge counting and does not depend on strain, the polarization charge is affected by strain through the piezoelectric response of the system. This means that even when $\lambda_{E}=0$ in equation~(\ref{eq:lambdaE}), we can still apply strain to restore an unbalanced polarization charge and thus a finite electric field.

\subsection*{Wannier functions of transition metal dichalcogenides}

\begin{figure}
\includegraphics{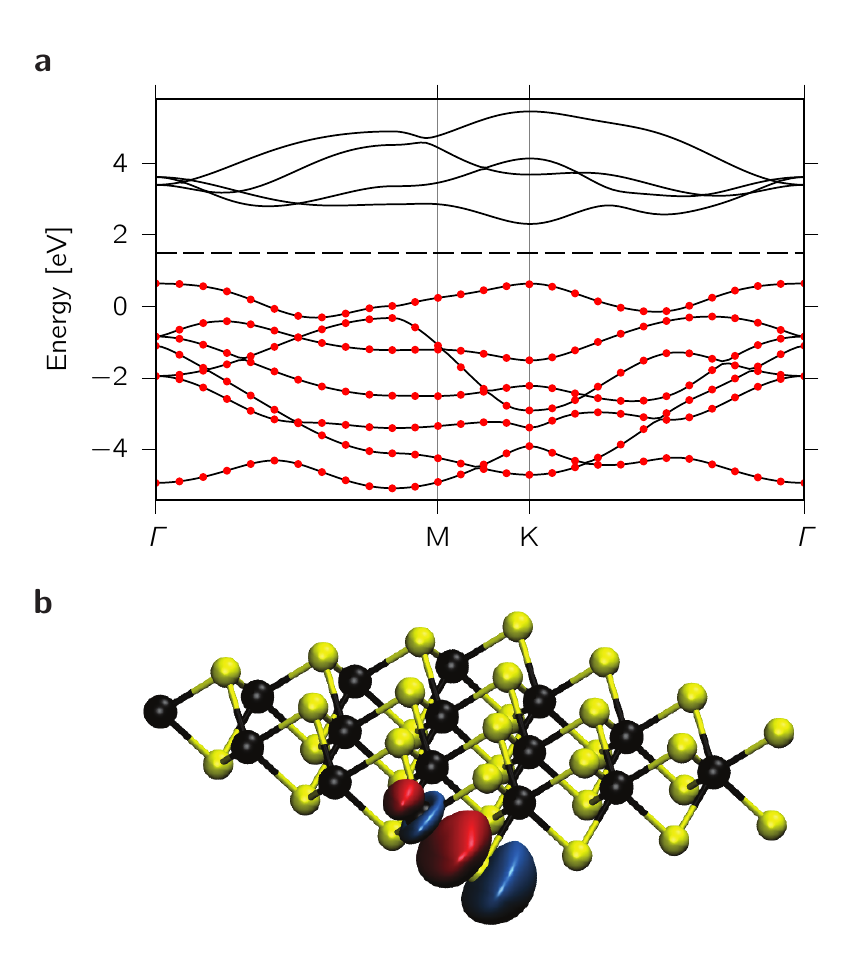}
\caption{\textbf{Band structure and Wannier functions of MoS$_{\bm{2}}$. a},
MoS$_{2}$ band structure along the path in reciprocal space connecting
the high-symmetry points $\Gamma$--M--K--$\Gamma$. Solid lines: original
valence and conduction bands generated directly from our DFT calculation.
The dashed line separates valence bands from conduction bands. Red
circles: Wannier-interpolated valence bands. The isolated top valence
band is associated with the Wannier function shown in Fig.~\ref{fig:wannier}d
of the main paper. The lower six valence bands can be mapped into maximally
localized Wannier functions that are symmetrically equivalent to the
one in panel\textbf{ b} and show strong hybridization between sulphur
\emph{p}-orbitals and molybdenum \emph{d}-orbitals.\label{fig:mos2_suppl}}
\end{figure}

In order to compute the bulk formal polarization of transition metal
dichalcogenides like MoS$_{2}$ we need the centers of the Wannier
functions associated with the valence bands. Including the deepest
electronic states into the ionic cores, we are left with six electrons
in the outermost \emph{d}-orbitals of the transition metal and four
\emph{p}-electrons for each chalcogen\cite{Leb2009}. These atomic
orbitals give rise to the eleven bands shown in Fig.~\ref{fig:mos2_suppl}a.
Seven of them are fully occupied (valence bands) and separated from
the lowest four conduction bands by a direct energy gap at the Brillouin
zone corners. Here and in the following we focus on the representative
case of MoS$_{2}$ although similar conclusions can be drawn for other
isoelectronic compounds like MoSe$_{2}$ or WS$_{2}$. A standard 
localization procedure~\cite{Mar2012} allows us to associate
the six lowest valence bands with as many Wannier functions that are
localized on sulphur atoms with lobes pointing towards one of the
nearest neighbors (see also Fig.~\ref{fig:wannier}c in the paper).
As shown in Fig.~\ref{fig:mos2_suppl}b, such Wannier functions arise
from the hybridization between \emph{p}-orbitals on sulphur and \emph{d}-orbitals
on molybdenum. Less trivial is the topmost valence band, which is
disentangled from the others. A careful analysis shows that this isolated
band can be mapped into a rather broad Wannier function with a dominant
\emph{d}-character, located at the center of the hexagonal unit cell
(see Fig.~\ref{fig:wannier}d in the main paper). As a measure
of the reliability of these Wannier functions we show in Fig.~\ref{fig:mos2_suppl}
that a Wannier interpolation procedure~\cite{Mar2012} is able to
reproduce precisely the first-principles valence band structure. In
addition, by applying equation~(3), it is possible
to prove that $m=2$ for MoS$_{2}$ (and similarly for all other group-VI
transition metal dichalcogenides).

\subsection*{Conformers of functionalized honeycomb lattices}

\begin{figure*}
\centering
\includegraphics[width=\linewidth]{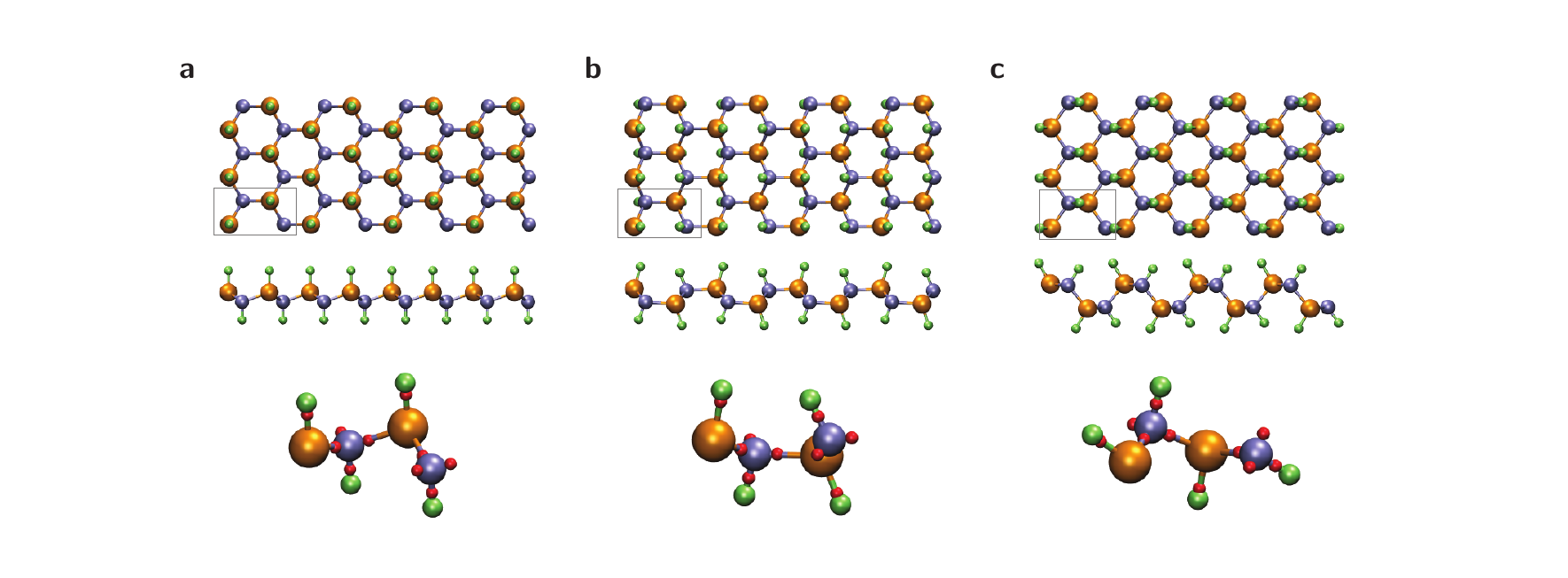}
\caption{\textbf{Possible conformers of functionalized honeycomb lattices}.
Top and lateral views of the three possible configurations of hydrogenated
boron nitride that are compatible with a eight-atom unit cell: chair
(\textbf{a)}, boat (\textbf{b}), and stirrup (\textbf{c}). A gray
rectangle highlights the eight-atom unit cell adopted in the calculations.
Bottom panels show a magnified perspective view of each unit cell
in which the Wannier function centers are also reported as small red
spheres. \label{fig:conformers}}
\end{figure*}

\begin{table}[t]
\caption{Ground-state energy ($E_{g}$) together with atomic ($E_{a}$) and molecular ($E_{m}$) formation
energies of hydrogenated and fluorinated BN in different configurations
(chair, boat, stirrup). For H functionalization the stirrup conformer
is the most stable, while in the case of fluorination the chair configuration
is preferred. A negative atomic formation energy in all cases suggests that
fluorination and hydrogenation should be achievable in atomic atmosphere.\label{tab:stability}}
\centering\begin{tabular}{ccccc}
\toprule 
 &  & $E_{g}$ (eV/atom) & $E_{a}$ (eV/atom) & $E_{m}$ (eV/atom)\tabularnewline
\midrule
\multirow{3}{*}{H} & chair & $-98.533$ & $-1.032$ & $0.102$\tabularnewline
 & boat & $-98.551$ & $-1.050$ & $0.084$\tabularnewline
 & stirrup & $-98.573$ & $-1.072$ & $0.062$\tabularnewline
\midrule
\multirow{3}{*}{F} & chair & $-427.471$ & $-1.354$ & $-0.645$\tabularnewline
 & boat & $-427.415$ & $-1.298$ & $-0.589$\tabularnewline
 & stirrup & $-427.441$ & $-1.324$ & $-0.615$\tabularnewline
\bottomrule
\end{tabular}
\end{table}

As we mentioned in the paper, selective functionalization of a parent
heteroatomic honeycomb crystal (like BN) introduces a polar discontinuity
in the system. We focused in particular on covalent hydrogenation
or fluorination, when H or F atoms are adsorbed on both sides of the
honeycomb lattice similarly to what happens in graphane and fluorographene.
Several configurations are possible, depending on the pattern formed
by the adsorbed atoms above and below the 2D sheet. Taking boron nitride
as parent honeycomb lattice, Fig.~\ref{fig:conformers} shows the
configurations compatible with an eight-atom unit cell: (a) chair,
(b) boat, and (c) stirrup\cite{Sof2009,Bha2011}. By projecting the
atomic positions in plane, we notice that their 2D spatial group (wallpaper
group) is different, being \emph{cm} for boat and stirrup isomers
while \emph{p3m1} for the chair structure. According to Ref.~\onlinecite{Jad2013}
the chair configuration is the only one compatible with a quantized
(topological) polarization [see equation~(2)].
For the other conformers, instead, symmetry only constrains one component
of the in-plane polarization vector, leaving the other one completely
undetermined. Even though in principle the chair configuration could
seem to be optimal since it ensures a stable (topological) polarization,
the strain arising from the lattice mismatch between pristine and
functionalized forms breaks the symmetries that protect quantization.
In addition, we shall see that all conformers support a different
polarization with respect to the parent material and thus give rise
to a polar discontinuity. For these reasons, all conformers give rise
to analogous physical effects, and only the chair configuration
has been discussed in the paper for the sake of
simplicity since it shares the same (2D) symmetry of the parent material
and thus its bulk polarization can be expressed through equation~(\ref{eq:pol_honey}). 

For the purpose of engineering a polar discontinuity, all configurations
are in principle equally relevant and one needs to assess their relative
stability. In table~\ref{tab:stability} we focus on boron nitride
and we report the ground-state energy together with atomic and molecular formation energies (per
atom) of the three configurations for both hydrogenated and fluorinated
structures. The atomic formation energy is defined as
\begin{equation}
E_{a}=E_{g}-E_{p}-N_{{\rm X}}E_{{\rm X}}\,,\label{eq:Ea}
\end{equation}
while the molecular formation energy reads
\begin{equation}
E_{m}=E_{g}-E_{p}-\frac{N_{{\rm X}}}{2}E_{{\rm X_{2}}}\,.\label{eq:Em}
\end{equation}
In equations~(\ref{eq:Ea}) and (\ref{eq:Em}), $E_{g}$ is
the total energy of functionalized BN in its optimized geometry, $E_{p}$
the total energy of the parent BN sheet, $N_{{\rm X}}$ the number
of H or F atoms, and $E_{{\rm X}}$ ($E_{{\rm X_{2}}}$) their atomic
(molecular) total energy. A negative atomic (molecular) formation energy reveals
if the functionalization is likely a favorable process in the presence
of the adsorbate in atomic (molecular) form. We immediately notice
that for hydrogenated BN the chair configuration not only has a positive
molecular formation energy, but it is even less stable than the boat and stirrup
isomers. This is in agreement with previous results on a similar level
of theory~\cite{Aver2009,Wang2010,Bhat2010,Sam2012}. On the contrary,
for fluorinated BN the repulsion between fluorine atoms favors a situation
in which neighboring F atoms are on opposite sides of the BN sheet.
As a consequence, the chair configuration is the most stable, with
both the atomic and molecular formation energies being negative. 
Since functionalization is typically performed in atomic atmosphere, 
our results suggest that both fluorination and hydrogenation are achievable for BN. 
Although these results based on equations~(\ref{eq:Ea}) and (\ref{eq:Em})
are valid only in the limit of zero temperature, we have verified that 
our conclusions do not change even at standard ambient temperature and pressure conditions.
Indeed, by replacing in equations~(\ref{eq:Ea}-\ref{eq:Em}) 
the total atomic or molecular energy with the chemical potential of the gas 
(including translational, rotational, vibrational, and nuclear contributions), 
the gain in formation (free) energy reported in table~\ref{tab:stability} decreases by less than 0.3~eV/atom.

We now want to verify that a polar discontinuity between pristine
and functionalized honeycomb lattices arises irrespectively of the
specific configuration (chair, boat, or stirrup) of the adsorbed atoms.
In order to compute the bulk formal polarization, we need the Wannier
function centers for each conformer. These are shown in Fig.~\ref{fig:conformers}
in the case of hydrogenated BN. We report results
only for hydrogen since the case of fluorine is completely analogous.
In fact, the six additional positive ionic charges of F with respect to H 
are completely balanced by three additional Wannier functions that are symmetrically arranged around the F atom.
We notice that the three configurations
are qualitatively very similar: In each unit cell we find a Wannier
function centered approximately mid-bond between each H and N or B
atoms and other three around each anion (N in this case) along directions
pointing towards its nearest B neighbors. For the chair conformer,
this is in agreement with the schematic picture in Fig.~\ref{fig:wannier}b.
Thus, only minor quantitative variations occur between the conformers,
while more radical qualitative features distinguish them from pristine
BN as already mentioned in the paper (see Figs.~\ref{fig:wannier}a
and b). In particular the formal polarization is always orthogonal
to the zigzag direction and reads (in units of $e/$Å): $-0.257$
(chair), $-0.288$ (boat), $-0.202$ (stirrup), and $-0.398$ (pristine
BN). Although piezoelectric contributions should be taken into account
when considering interfaces, this analysis already shows that for
any conformer selective hydrogenation leads to a polar discontinuity
in BN honeycomb lattices. Indeed, we have verified (through a more
careful simulation of a zigzag interface between pristine and functionalized
BN) that free charges appear at the boundaries as a result of the
polarization-induced electric fields independently of the specific
arrangement (chair, boat, or stirrup) of hydrogen atoms.

\end{document}